\documentclass{article}
\usepackage{ijcai22}

\usepackage{times}

\usepackage{soul}
\usepackage{url}
\usepackage[hidelinks]{hyperref}
\usepackage[utf8]{inputenc}
\usepackage[small]{caption}
\usepackage{graphicx}
\usepackage{amsmath}
\usepackage{booktabs}
\usepackage{kotex}

\usepackage{tikz}
\usepackage{amsmath}
\usepackage{mathtools}
\DeclareMathOperator*{\argmax}{arg\,max}
\DeclareMathOperator*{\argmin}{arg\,min}

\usepackage{booktabs}       % professional-quality tables
\usepackage{amsfonts}       % blackboard math symbols
\usepackage{microtype}
\usepackage{booktabs}       % professional-quality tables

\usepackage{subfig}
\usepackage{wrapfig}
\usepackage{booktabs, multirow} % for borders and merged ranges

\usepackage{filecontents}

\usepackage{bbding}
\usepackage{pifont}
\usepackage{wasysym}
\usepackage{algorithm}
\usepackage{algorithmic}
\usepackage[toc,page]{appendix}

\iftrue
\title{Online Evasion Attacks on Recurrent Models:\\
The Power of Hallucinating the Future}

\author{
Byunggill Joe$^1$\and
Insik Shin$^{1}$\And
Jihun Hamm$^2$\footnote{Corresponding author.}\\
\affiliations
$^1$ School of Computing, KAIST, Daejeon, South Korea\\
$^2$ Department of Computer Science, Tulane University, Louisiana, USA\\
\emails
\{byunggill.joe, insik.shin\}@kaist.ac.kr,
{jhamm3}@tulane.edu
}
\fi

\begin{document}

\maketitle

\begin{abstract}
%It is important to comprehensively assess the vulnerabilities of recurrent models in online tasks such as autonomous driving, where security is critical.
Recurrent models are frequently being used in online tasks such as autonomous driving, and a comprehensive study of their vulnerability is called for.
Existing research is limited in generality only addressing application-specific vulnerability or making implausible assumptions such as the knowledge of future input. In this paper, we present a general attack framework for online tasks incorporating the unique constraints of the online setting different from offline tasks.
Our framework is versatile in that it covers time-varying adversarial objectives and various optimization constraints, allowing for a comprehensive study of robustness.
Using the framework, we also present a novel white-box attack called Predictive Attack that `hallucinates' the future. The attack achieves 98 percent of the performance of the ideal but infeasible clairvoyant attack on average. We validate the effectiveness of the proposed framework and attacks through various experiments. 

\end{abstract}

\section{Introduction}
% Page instruction.
% \textcolor{red}{
% \begin{itemize}
% \item Paper submissions must be at most 7 pages in double-column ACM format,
% \item reference 2 pages.
% \end{itemize}
% }
Deep neural networks (DNN) are discovered to be surprisingly vulnerable to imperceptibly small input noises~\cite{szegedy,fgsm}. Many different types of vulnerabilities of DNNs have been demonstrated in various tasks, including classification, regression, and generative methods~\cite{pgd,attack-vae,autoregressive_model_attack}. Most attacks have focused on offline evasion attacks on non-temporal models, such as the image classification model, where an attacker has access to the whole input example, such as an image.
% Research in deep neural networks (DNNs) has made tremendous progress in the past decade, and DNNs are achieving human-level performances in various tasks~\cite{imagenetcnn,resnet,nlpunifiedarch,adversarial_reinforcement}. However, they are also shown to be surprisingly vulnerable to imperceptibly small input noises. Since the first report of the adversarial attack~\cite{szegedy}, many different types of vulnerabilities of DNNs have been demonstrated in various tasks, including classification, regression, reinforcement learning, and generative methods~\cite{fgsm,pgd,attack-vae,autoregressive_model_attack}.  Most of the adversarial attacks have focused on offline evasion attacks on non-temporal models, such as the image classification model where an attacker has access to the entire input.

However, there are many security-critical tasks that involve recurrent neural networks (RNN) performed in an online fashion~\cite{agrometeorological_forecasting,electricity_consumption_prediction,realtime_object_tracking}. For instance,  mortality prediction ~\cite{mimic_mortality} continuously monitors hospitalized patients for early warning, and an autonomous driving agent that uses sensors to decide the steering angle of the vehicle \cite{autonomous_driving_survey}. Unlike the offline setting, an attacker cannot observe the entire input sequence because inputs arrive as a real-time stream to a victim and an attacker.

\begin{figure}[t]
\centering
\includegraphics[width=1.0\linewidth]{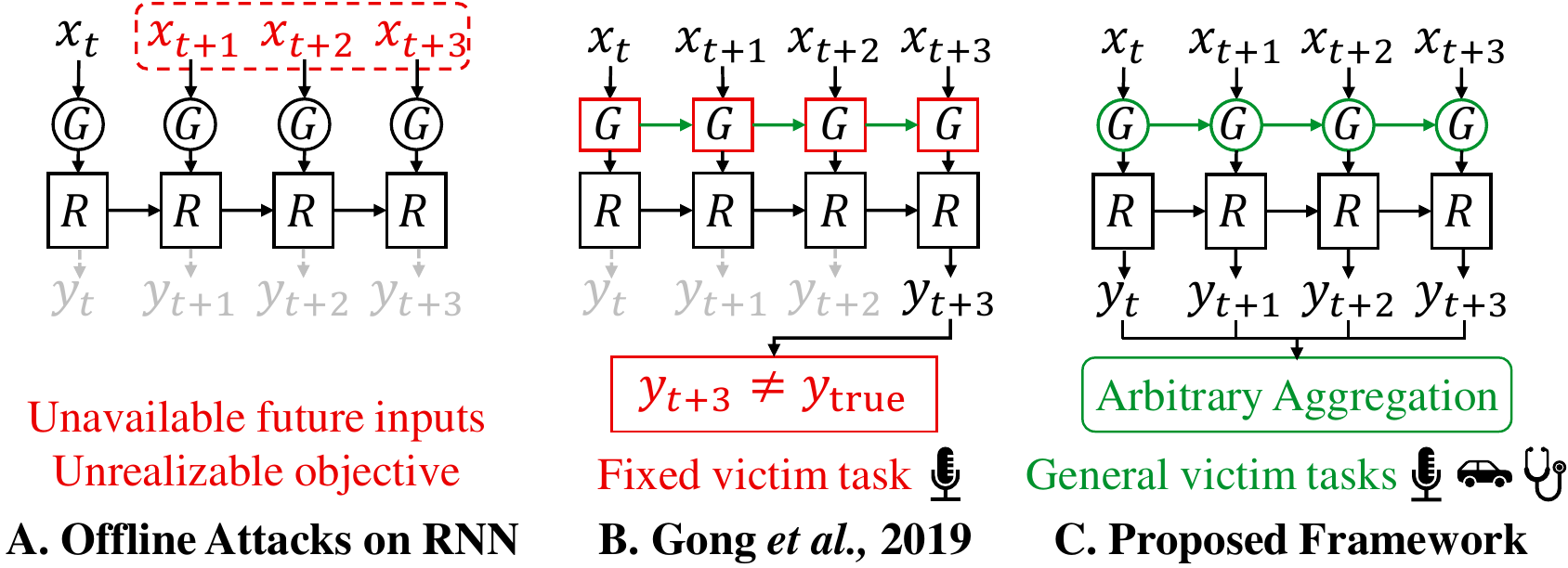}
  \caption{Framework comparison for attacking an RNN of an online task. 
  ``G'': Attack perturbation generator, ``R'': Victim RNN.}
\label{fig:framework_differentiation}
  %\vspace{-1.5em}
\end{figure}

%~\cite{intrusion_detection_attack,realtime_universal_speaker_recognition_attack,appending_adversarial_attack,time_series_classification_attack,autoregressive_model_attack,time_series_elastic_attack,time_series_regression_attack}
Previous works~\cite{realtime_universal_speaker_recognition_attack,time_series_classification_attack,autoregressive_model_attack,time_series_elastic_attack} evaluated vulnerabilities of RNNs but they implicitly assumed that future inputs are observable, which is implausible (\autoref{fig:framework_differentiation}-A, dashed box).
Meanwhile, \cite{realtime_adversarial_attack} introduced a framework for real-time attack (\autoref{fig:framework_differentiation}-B) with two constraints unique to the online problem: 1) future inputs are unobservable, and 2) the past inputs are unchangeable. 
However, the framework is rather specific to speech recognition (\autoref{fig:framework_differentiation}-B, red boxes),
where classification is performed only once after receiving the entire speech. Such an approach is inapplicable to dynamic online tasks such as autonomous driving,
where the agent has to continuously decide the steering angle. 

In this paper, we propose a more general framework of online evasion attacks\footnote{\url{https://github.com/byunggilljoe/rnn_online_evasion_attack}. Appendix is included.} allowing a victim recurrent model to make continuous predictions or decisions (\autoref{fig:framework_differentiation}-C, circled G and a round box).
Our framework can accommodate various adversarial objectives on RNN to address different attack scenarios. 
In particular, our framework makes time-varying adversarial objectives possible, unlike previous approaches to attack at a specific time. The versatility of our framework will allow for a comprehensive robustness study of recurrent models.
To showcase of the versatility, we reformulate the objective of real-time attack~\cite{realtime_adversarial_attack}, and present novel adversarial objectives such as Time-window and Surprise objectives using our framework.

\begin{figure}[t]
\centering
\includegraphics[width=0.9\linewidth]{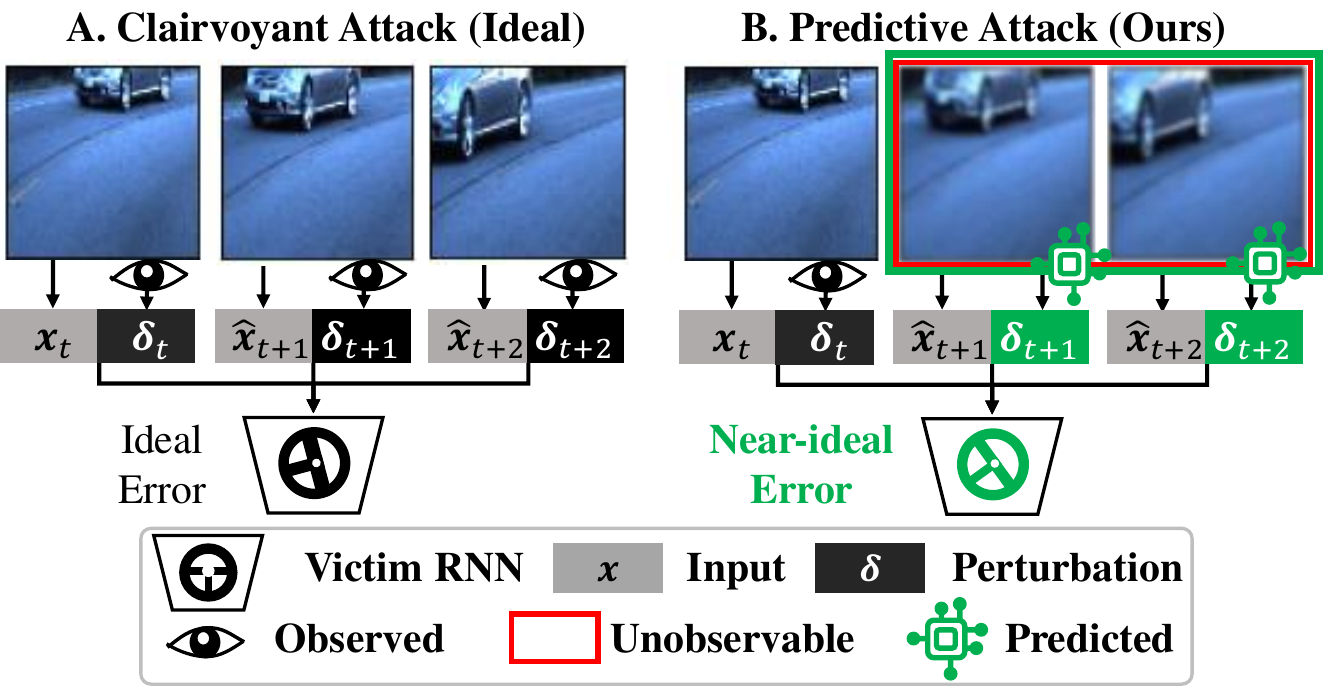}
%\vspace{-0.7em}
  \caption{Attacking a vision-based autonomous driving agent to change the steering angles.
  Clairvoyant Attack (A) can see all future inputs and achieves the best attack but it is unrealizable. Predictive Attack (B) emulates the clairvoyant by hallucinating the future using a predictive model of the input sequence.
  }
\label{fig:overview}
  %\vspace{-0.8em}
\end{figure}

%\cite{fgsm,pgd,fab,square}.
As an effective solution to our framework, we propose a novel white-box attack called Predictive Attack (Figure~\ref{fig:overview}-B). An ideal solution to the online problem is the clairvoyant attack (Figure~\ref{fig:overview}-A), where an attacker does not suffer from the online constraints, foreseeing the entire future input. Thus the clairvoyant attack can find attack perturbations with existing offline methods~\cite{pgd,fab}.
Instead of clairvoyance, Predictive Attack `hallucinate' the future (Figure~\ref{fig:overview}-B, green box) with a trained predictive model of input sequences, mimicking the crystal ball of a clairvoyant. Since accurate prediction can be difficult, we propose an additional alternative attack called IID Attack that replaces accurate prediction with IID sampling. They perform surprisingly well, and we ascribe this to the importance of considering the hidden states and input orders when attacking recurrent models.

% Based on our framework, we propose a novel white-box attack called Predictive Attack. A naive solution to the general online attack problem is the greedy approach (Figure~\ref{fig:overview}. A) that only considers the effect of the current perturbation on current output, ignoring the future; therefore, it results in poor attack performance. 
% In contrast, an ideal solution to the online problem is the clairvoyant attack (Figure~\ref{fig:overview}. B) in which the attacker can foresee the entire future input. Under this unrealistic assumption, the online attack problem becomes equivalent to an offline attack problem whose solution can be found by existing methods~\cite{fgsm,pgd,fab,square}.
% Instead of clairvoyance, we propose to `hallucinate' the future with a trained predictive model of input sequences, replacing the crystal ball of a clairvoyant. We first present Predictive Attack (Figure~\ref{fig:overview}. C) that considers the effect of the current perturbation on future outputs with a predictive model in the white-box setting. 
% Since accurate prediction can be difficult, we propose two additional alternative attacks called IID Attack and Random Attack, which respectively replace accurate prediction with IID sampling and Random noise sampling. They perform surprisingly well and better than the naive greedy attack by a large margin. We ascribe this to the importance of considering the hidden states when attacking recurrent models.

We evaluate our attacks using six datasets. Our predictive attack approaches 98\% of the performance of the clairvoyant on average.
We perform further empirical analysis of the predictive attacks demonstrating the versatility of our framework and attack robustness.
We summarize our contributions as follows.
%  We evaluate our attacks using MNIST~\cite{mnist}, FashionMNIST~\cite{fashion-mnist}, mortality prediction~\cite{mimic_mortality}, and autonomous driving~\cite{udacity}. Our predictive attack achieves 1.21$\sim$10.6 times the performance of the greedy attack that ignores input temporal dependence and victim model dynamics, and approaches closely the performance of the clairvoyant attack (Figure~\ref{fig:performance-evaluation-predictive}). 
% We perform further empirical analysis of the predictive attacks, including the impacts of different conditions, comparison of the three proposed attacks, and the effectiveness of new online attack objectives. 
% We summarize our contributions as follows.
\begin{itemize}
\item We introduce a general formulation of online evasion attacks on recurrent models, which can accommodate various types of attack objectives and constraints, allowing for a comprehensive study of robustness.
\item We propose two novel white-box attacks, Predictive Attack and IID Attack, based on hallucination of the future to emulate the ideal clairvoyant attacker.   
\item We evaluate the performance of our attacks under various conditions using real-world data and demonstrate the versatility and robustness of our framework and attacks. 
\end{itemize}

%\BG{too repetitive?}

% ----
% \begin{itemize}
%     \item Machine learning은 여러 분야에서 좋은 성능을 보인다.
%     \item 하지만 adversarial attack에 의해 그 성능이 robust하지 않다는 것이 보여졌다.
%     \item 하지만 이러한 adversarial attack은 많은 경우 image classification 도메인에 한정되어 연구되었다.
%     \item 이 논문에서는 sequential 모델에 대한 adversarial attack, online evasion attack을 다룬다. \JH{Attack on real-time classification vs attack on sequential models ... I'm not sure which is better. But sequential models can be attacked offline as well.}
%     \item \JH{Key question: What can online evasion attack do that other attacks cannot do? Why is it important?}
%     \item \JH{Real-time (or sequential?) classification의 중요성 -- patient monitoring의 예와 autonomous driving agent의 예.}
%     \item 실제 산업에서는 sequential 모델도 많이 사용되고 있지만 아직 이 종류의 모델에 대한 adversarial attack 연구는 많이 되어 있지 않다.
%     \item 우리는 기존 image classifier에 대한 adversarial attack을 sequential model로 확장하는데 한계가 있다는 것을 보인다. 
%     \item 이러한 한계를 극복하기 위해 predictive model을 사용한 sequential model에 대한 새로운 attack을 제시하고 실험을 통해 greedy한 approach보다 많은 성능 향상을 가져옴을 보인다.
%     \item \JH{기존 real-time attack과의 차이.} 
%     \item \JH{기존 online-learning attack과의 차이.}
% \end{itemize}

% Contribution
% \begin{itemize}
%     \item 기존 adversarial evasion attack의 sequential model로의 확장 한계 발견.
%     \item 이를 해결하기위해 미래 입력 예측 모델을 활용한 online adversarial evasion attack을 처음으로 제시.
%     \item 실험으로 구현하여 공격의 effectiveness를 보임.
% \end{itemize}

%\section{Related Work}
%\input{background_relatedwork}

\section{Setting}
\paragraph{Inputs and Outputs.}
An input stream/sequence $\boldsymbol{x}$ of length $L$ is a sequence of $n$-dimensional vectors
$(x_1, x_2, ..., x_L)\in \mathbb{R}^{L\times n}$ where the index refers to time step. 
Similarly, the output sequence $\boldsymbol{y}$ of length $L$ is sequence of outputs $(y_1, y_2, ..., y_L)$ where $y_i\in\mathbb{R}$ for a regression problem and $y_i \in \{1,...,C\}$ for a classification problem.
%The paired sequence $\boldsymbol{z}$ of size $L$ is a sequence of input-output pairs $((x_1,y_1), ..., (x_L,y_L))$.
%\begin{gather}
%\boldsymbol{x}=(x_{1}, x_{2}, ..., x_{l}), x_{i}\in \mathbb{R}^{n}.\\
%\boldsymbol{y} = (y_1,...,y_{l}), y_i\in \mathbb{R}.
%\end{gather}
%A labeled dataset is the collection of paired sequences 
%$\mathcal{D}=\{\boldsymbol{z}^1, ..., \boldsymbol{z}^N\}$.%\{({x_i}, {y_i})_{i=1}^l\
%\BG{Do we need to mention transferability and our attack is grey box?}
\paragraph{Victim Task.}
%We consider tasks that use input sequences to continuously predict output sequences. % to  temporal dependency in inputs or outputs. Temporal dependency means that current data (input or output) have a likelihood function conditioned on past data. Thus, a dataset $\mathcal{D}$ of a task is a set of sequences $\mathcal{D}=\{({x_i}, {y_i})_{i=1}^l\}$ that encode temporal dependency. 
We attack recurrent neural networks (RNN) that continuously predict the output at each time step. %An RNN models the temporal dependency within/between the input sequence and the output sequence and uses the hidden state to summarize the past input. 
Formally, an RNN is a pair of functions $f_\theta$ and $g_\theta$. At time $t$, $f_\theta:\mathbb{R}^n\times\mathbb{R}^m \rightarrow \mathbb{R}$ predicts the current output by $y_t=f_\theta(x_t,h_t)$ using the current input $x_t$ and the hidden state $h_{t}\in\mathbb{R}^m$. The dynamics of the RNN is determined by $g_\theta:\mathbb{R}^n\times\mathbb{R}^m \rightarrow \mathbb{R}^m$ which maps ($x_t$, $h_t$) to the next hidden state by $h_{t+1} = g_{\theta}(x_{t},h_{t})$.
%An RNN $(f_\theta, g_\theta)$ is typically trained to minimize the loss $\mathcal{L}$ such as cross-entropy or mean squared error averaged over the sequence.
\paragraph{Threat Model.}
% \begin{itemize}
%     \item what an attacker can do. (whitebox assumption)
%     \item Need to be short.
% \end{itemize}
We assume attackers have white-box access to a victim model. Attackers can define an attack objective and loss and  can compute the derivative of the loss with respect to an input. Also, Attackers have access to some examples of input streams.

\section{Online Evasion Attack Framework}
\paragraph{Problem.}
The attacker aims to mislead a victim RNN model $(f_\theta,g_\theta)$ to output the (adversarial) target labels or values $(y^{a}_1,\cdots,y^{a}_L)$
by using the perturbed input sequence $(x_1+\delta_1, \cdots, x_L + \delta_L)$. 
This is done by minimizing\footnote{The current description is for targeted attacks but we can also perform untargeted attacks as well.} the aggregate value of the losses $\mathcal{L}^{\text{adv}}_1,\cdots,\mathcal{L}^{\text{adv}}_L$:%at each time and $\mathrm{Agg}(\cdot)$ refers to a method of temporal aggregation,
%as follows: %$\sum^{l}_{i=1} \mathcal{L}_{\text{adv}}(x_{i},y_{i},\delta_{i},h^{\delta}_{i})$, (Eqn. \ref{eq:problem_definition_attack_loss}).
\begin{gather}
\boldsymbol{\delta}=\argmin_{\boldsymbol{\delta}=(\delta_1,\cdots,\delta_L) \in \Delta}\; 
\mathrm{Agg}
\left(\mathcal{L}^{\text{adv}}_1,\cdots,\mathcal{L}^{\text{adv}}_L\right),\;\mathrm{where} \label{eq:problem_definition_attack_loss_general}
\end{gather}
$\mathcal{L}^{\text{adv}}_i$ is the loss at time $i$: $\mathcal{L}^{\text{adv}}_i=\mathcal{L}(f_{\theta}(x_{i}+\delta_{i}, h^{\delta}_{i}), y^{a}_{i})$, and
$h^{\delta}_i$ is the hidden state of the RNN at time $i$:
%\mathcal{L}^{\text{adv}}_i=\mathcal{L}^{\text{adv}}(x_{i},y^{a}_{i},\delta_{i},h^{\delta}_{i})=\mathcal{L}(f_{\theta}(x_{i}+\delta_{i}, h^{\delta}_{i}), y^{a}_{i}),\\
\begin{equation}
h^{\delta}_i = g_{\theta}(x_{i-1}+\delta_{i-1},h^{\delta}_{i-1}). \label{eq:problem_definition_h_update}
\end{equation}
The $\mathrm{Agg}(\cdot)$ refers to a method of temporal aggregation,
and $\Delta$ refers to any constraint on the perturbation sequence.
%Due to the dynamics (Eqn. \ref{eq:problem_definition_h_update}) of the victim, note that the current perturbation $\delta_{t}$ affects the future hidden state $h^{\delta}_{i}$ for $i=t+1,...,L$ and therefore the future losses too. 
Compared to previous work~\cite{realtime_adversarial_attack}, the present formulation (\autoref{eq:problem_definition_attack_loss_general}$\sim$\ref{eq:problem_definition_h_update}) is much more flexible since the loss and the target are allowed to be time-varying. % with all inputs and outputs.
%To show the flexibility, we present two examples possible in our framework in Method section.
For concreteness, we will use the temporal summation $\mathrm{Agg}(\mathcal{L}_1,\cdots,\mathcal{L}_l) = \sum_{i=1}^L \mathcal{L}_i$ and 
the $\ell_p$ constraint $\Delta=\{\|\delta_i\|_p \leq \epsilon,\;\;\forall i\}$ by default:
\begin{gather}
\boldsymbol{\delta}=\argmin_{\|{\delta_i}\|_{p} \leq \epsilon,\;\forall i}\; \sum^{L}_{i=1} \mathcal{L}^{\text{adv}}(x_{i},y^{a}_{i},\delta_{i},h^{\delta}_{i})\label{eq:problem_definition_attack_loss}.
%\mathcal{L}_{\text{adv}}(x_{i},y_{i},\delta_{i},h^{\delta}_{i})=\mathcal{L}(f_{\theta}(x_{i}+\delta_{i}, h^{\delta}_{i}), y_{i}),\\
%h^{\delta}_i = z_{\theta}(x_{i-1}+\delta_{i-1},h^{\delta}_{i-1}). \label{eq:problem_definition_h_update}
%|\boldsymbol{\delta}|_p \leq \epsilon.
%\sum_{(\boldsymbol{x},\boldsymbol{y})\in\mathcal{D}}
\end{gather}
\paragraph{Online Constraints.}
Critically different from the much-studied offline attacks, an online attack has to follow physical constraints~\cite{realtime_adversarial_attack}. 
Firstly, an attacker cannot perturb the future or the past input but only the current input. Therefore, to solve \autoref{eq:problem_definition_attack_loss} an attacker has to solve
\begin{equation}\label{eq:online_attack_loss_orig}
\delta_{t}=\argmin_{\|\delta_{t}\|_p\leq \epsilon}\;
\sum_{i=t}^L \mathcal{L}^{\text{adv}}(x_{i},y^{a}_{i},\delta_{i},h^{\delta}_{i})
\end{equation}
at each time step $t=1,\cdots, L$, which is the core of the general online attack. Since the losses of the past ($i<t$) are unchangeable they do not appear in the sum of the losses.
An important thing to note is that the current perturbation $\delta_t$ affects all future losses $\mathcal{L}^{\text{adv}}_{t+1},\mathcal{L}^{\text{adv}}_{t+2},\cdots$ due to the nature of RNNs, which we call \textbf{victim model dynamics} property. A successful online attack therefore has to exploit this property.
Furthermore, the sum can be rewritten as
\begin{gather}
%\text{For each}\;t=1, \cdots, L,\;\;\text{find the current perturbation}\\
\delta_{t}=\argmin_{\|\delta_{t}\|_p\leq \epsilon}\;\underbrace{\mathcal{L}^{\text{adv}}(x_{t},y^{a}_{t},\delta_{t},h^{\delta}_{t})}_{\mathclap{\text{(A) Current}}} + \sum^{L}_{i=t+1}
\underbrace{\mathcal{L}^{\text{adv}}(x_{i},y^{a}_{i},\delta_{i},h^{\delta}_{i})}_{\mathclap{\text{(B) Future, \textcolor{red}{not observed}.}}}. \label{eq:online_attack_loss}%\\
%
%\text{such that, }||\boldsymbol{\delta}||_{p} \leq \epsilon.
\end{gather}
As the equation shows, this optimization problem cannot be solved directly due to the second constraint of the online attack: we do not know the future inputs $x_{t+1},x_{t+2},\cdots$.
%The objective function has two parts - one that depends on the current observable input (A) and one that depends on the unobservable future (B). 
%It is also apparent from the equation that the attack can only choose the current perturbation $\delta_i$.
Although seemingly impossible, we make it possible  by exploiting the \textbf{temporal dependence} of the inputs in a stream, on which the decisions of RNN depend ultimately. (Details in the next section.) 
To mount a successful attack, an online attacker has to exploit both victim model dynamics and temporal dependence. 

%In the online setting~\cite{realtime_adversarial_attack}, such as an autonomous driving model, it becomes harder to attack a victim since an attacker can not observe and manipulate an entire input stream $(x_1, x_2, ..., x_L)$ at once. In particular, an attacker 1) cannot observe the future input thus can not consider the victim dynamics and has limitations in using the temporal dependence. Also, the attacker 2) cannot change the past perturbations already delivered to the victim. In addition, our framework allows 3) the victim model to perform predictions at every step, so the attacker has to consider multiple outputs for a successful attack. 

\begin{figure}
\centering
\includegraphics[width=0.95\linewidth]{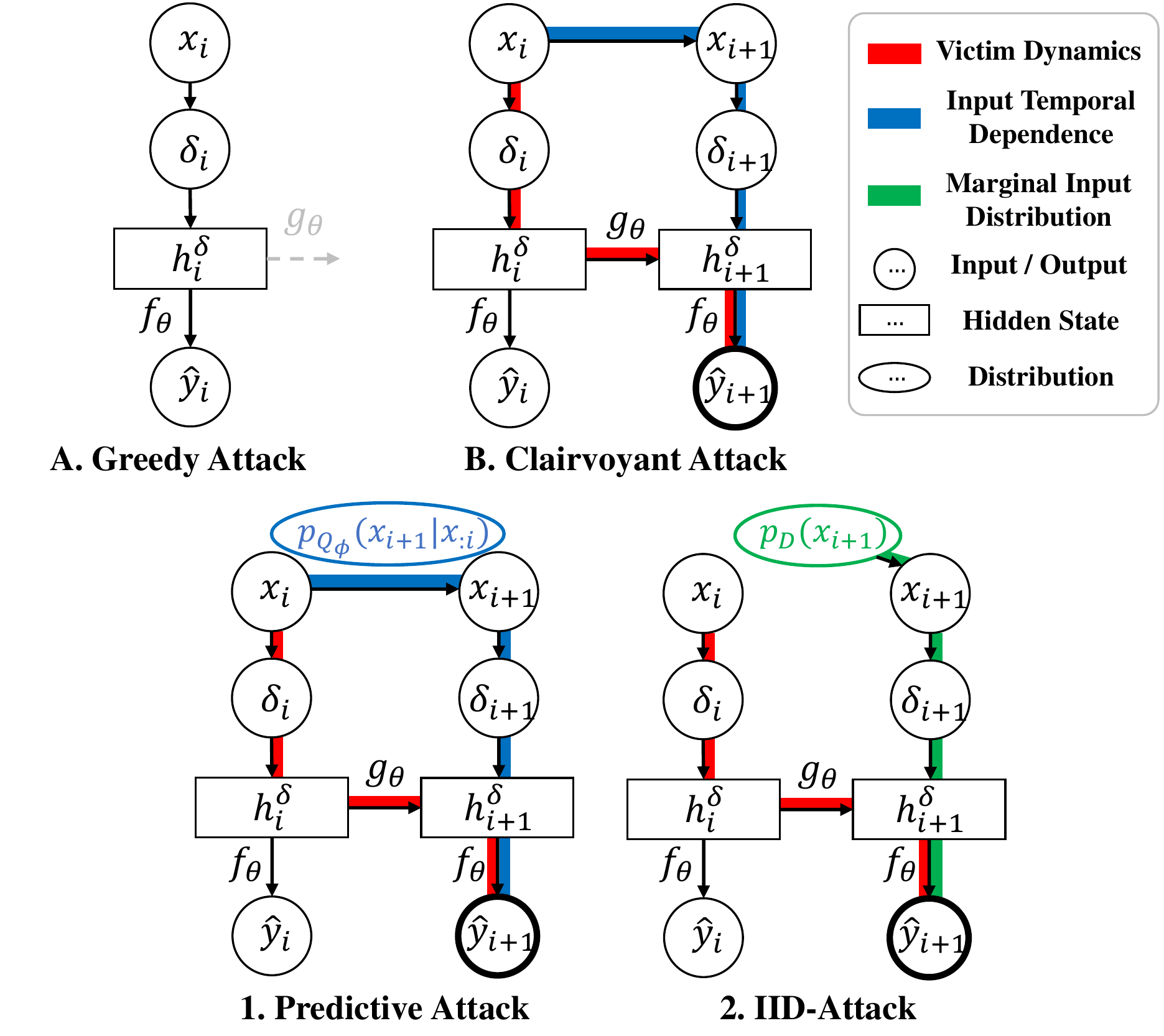}
%\vspace{-0.5em}
  \caption{
  Comparison of attack methods: reference (A and B) and proposed (1, 2). Please refer to Method for details.
   %\BG{I fixed it and changed ``Trained Input Distribution''$\to$``Input Distribution w/o Dependence'' $\to$``Marginal Input Distribution'', 
   %change f g h}
  }
  \label{fig:exploitable-components}
  %\vspace{-1.5em}
\end{figure}
\subsection{Greedy and Clairvoyant Attack}
We propose two reference attacks that exemplify a crude attack and an ideal attack. These attacks can help us understand the other attacks in the following sections. \textbf{Greedy Attack}~(\autoref{fig:exploitable-components}-A) provides a lower bound of attack performance which does not consider the victim model dynamics and the temporal dependency. This attack only considers the current loss (A) of \autoref{eq:online_attack_loss}.
 \textbf{Clairvoyant Attack}~(\autoref{fig:exploitable-components}-B) is an ideal, unrealizable attack that assumes the full observability of the future part of an input sequence; thus can fully use the victim model dynamics (\autoref{fig:exploitable-components}-B, red line), and temporal dependence (\autoref{fig:exploitable-components}-B, blue line). 
The Clairvoyant provides the upper bound of performance an attack can achieve.

\section{Method}

\subsection{Attacks using Future Hallucination}
Since the ideal Clairvoyant Attack is impossible, we replace the true future with a `hallucination' of it.
We propose two methods for the hallucination: using a predictive recurrent model (Predictive Attack), and random data substitution (IID Attack).
\paragraph{ Predictive Attack.}
Predictive Attack (Figure \ref{fig:exploitable-components}-1) uses a future predictive model to mimic Clairvoyant Attack.
%exploit the temporal dependence and victim model dynamics.
We define the attack objective of Predictive Attack as follows:
\begin{gather}
\begin{aligned}
\delta_{t}=&\argmin_{\|\delta_{t}\|_p\leq \epsilon}\;\underbrace{\mathcal{L}^{\text{adv}}(x_{t},y^{a}_{t},\delta_{t},h^{\delta}_{t})}_{\mathclap{\text{(A) Current}}}\\
& + 
\underbrace{\text{E}_{p(x_{t+1:}|x_{:t})}\left[\sum^{t+K}_{i=t+1}\mathcal{L}^{\text{adv}}(x_{i},y^{a}_{i},\delta_{i},h^{\delta}_{i})\right]}_{\mathclap{\text{(B) Future, \textcolor{blue}{$x_{t+1:}$ depends on $x_{:t}$}.}}}. \label{eq:predictive_attack_loss_full}%\\
\end{aligned}
\end{gather}
Due to linearity the second term can be simplified as
\begin{gather}
\sum^{t+K}_{i=t+1}
\text{E}_{p(x_{i}|x_{:t})}\left[\mathcal{L}^{\text{adv}}(x_{i},y^{a}_{i},\delta_{i},h^{\delta}_{i})\right].\label{eq:predictive_attack_loss}%
\end{gather}
Instead of directly modeling the distribution $p(x_{t+1:}|x_{:t})$, we undertake the easier task of generating the future input with a (stochastic) generative model $Q_{\phi}(x_{t+1}|x_{:t})$ that predicts the next input $x_{t+1}$ given $x_{:t}$ (\autoref{fig:exploitable-components}-1, blue). We restrict the number of the prediction steps to $K$, called {\bf lookahead} since we cannot consider all future inputs with finite resources.  

%Note that if the true distribution $p$ is known and can be used with Predictive Attack, it will perform similar to Clairvoyant Attack but not quite the same. This is because  there is uncertainty due to the inherent randomness of the future input. Therefore Predictive Attack based on the expected loss (Eqn.~\ref{eq:predictive_attack_loss_full}) using a known $p$ is an ideal and {\it realizable} attack, whereas Clairvoyant Attack is an ideal but {\it unrealizable} attack. 
We use another RNN to model the generator $Q_{\phi}$ to predict the next input $x_{t+1}$ from $x_{:t}$, using examples of input sequences as a training dataset. (Model details are in Experiments and
Appendix B.)
%\autoref{appendix-model-parameters}.)
%Note that $E_{Q_{\phi}}[\cdot]$ is computed by Monte-Carlo simulation~\cite{montecarlo_simulation} and uses the average over multiple sequences. For relatively short sequences where the future inputs are easier to  predict (e.g., the camera view of the road after 1 second as opposed to after 1 minute), the average can be approximated by just a single sequence with sufficient accuracy as we demonstrate in the experiments.
%\paragraph{Algorithm.}

\begin{algorithm}[t]
\begin{algorithmic}[1]
    \STATE{$ count \gets 0 $} 
    \WHILE{$ count < \text{MAX\_COUNT} $}
        %\STATE{$\hat{x}_{i+1}, ..., \hat{x}_{i+k} \gets Q_{\phi}(x_{1:i}), ..., Q_{\phi}(x_{1:i+k-1})$}\\
        \STATE{$\mathcal{L}_{\text{total}}(\delta_t) \gets \mathcal{L}_{\text{adv}}(x_{t},y^{a}_{t},\delta_{t},h^{\delta}_{t})\newline
        \textcolor{white}{.}\;\;\;\;\;\;\;+ \sum^{t+K}_{i=t+1}\text{E}_{Q_{\phi}(x_i|x_{:i-1})}[\mathcal{L}_{\text{adv}}(x_{i},y^{a}_{i},\delta_{i},h^{\delta}_{i})]$\newline
        \textcolor{white}{.}\;\;\;\;\;\;\;\; using Monte-Carlo to compute $E_{Q_{\phi}}[\cdot]$.}
        \STATE{$\forall i \in [t, t+K],$}
        \STATE{$\delta_i \xleftarrow{} \Pi_{\|{\delta_i}\|_p \leq \epsilon}[\delta_{i} - \alpha \text{sign}(\nabla_{\delta_i}\mathcal{L}_{\text{total}}(\delta_t)]$}
        \STATE{$\delta_i \xleftarrow{} \text{clip}(x_{i}+\delta_i) - x_{i}$}\newline
        \textcolor{white}{.}\;\;\;\;\;\;\;\;\text{It forces a valid range of perturbed inputs.}
        \STATE{$count \xleftarrow{} count + 1$}
    \ENDWHILE
    \RETURN{$\delta_{t}$}
\end{algorithmic}
  \caption{Predictive Attack at time $t$.}
    \label{alg:predictive-attack-algorithm}
\end{algorithm}

Algorithm \ref{alg:predictive-attack-algorithm} describes Predictive Attack's update rule for $\delta_{t}$, which is a variant of \cite{pgd}. The hyper-parameters $\text{MAX\_COUNT}$ and $\alpha$ determine the number of updates and the step size of an update. We elaborate more on this in Appendix A. %\autoref{appendix:algorithm-detail}.
\paragraph{IID Attack.}
Hallucinating the future based on an accurate $Q_\phi$ can be difficult due to the test-time cost of the prediction or the training-time cost of $Q_\phi$.
To relieve this, we present a heuristic, IID Attack, to replace the prediction model.
IID Attack (Figure \ref{fig:exploitable-components}-2) simply ignores the temporal dependence and predicts the future using IID sampling of the input data (Figure \ref{fig:exploitable-components}-2, green), that is, using $E_{p(x_{i})}[\cdot]$ instead of $E_{p(x_{i}|x_{:t})}[\cdot]$ in~\autoref{eq:predictive_attack_loss_full}.
Practically, this can be done by collecting a sufficient number of past input data and randomly choosing one of them as an IID example. 
Even with the incorrect prediction of IID, it is still using the victim model dynamics (Figure \ref{fig:exploitable-components}-2, red). Such a consideration makes a big difference compared to the current-only greedy perturbation $\delta_t$ as we will see.
% {\it So why is IID Attack any better than Greedy Attack?} Although IID is not hallucinating the future as accurately as Predictive Attack, it is still using the dynamics of the victim RNN (red lines) by estimating the impact of the current perturbation $\delta_t$ on the hidden state $h_t$ and consequently all the future outputs $y_{t+1:}$.
% Such a consideration of the future impact can make a big difference compared to  the current-only greedy perturbation $\delta_t$, which may set the hidden state $h_t$ less prone to be attacked in the future as the input changes over time. We validate this claim through experiments.
%\subsection{Online Attack Objectives with Temporal Specificity}
\begin{table}[t]
    \centering

    %\vspace{-0.8\baselineskip}
    %\scriptsize
    \footnotesize
    %\small
    %\begin{adjustbox}{minipage=\linewidth,scale=1.0}
      \centering
      {
\begin{tabular}{c| c r r c}
\toprule
    &&    &  & \textbf{Victim Clean}            \\
    \textbf{Dataset} &   \textbf{Task}  &   \textbf{$n$} & \textbf{$L$} & \textbf{Performance} \\
\midrule
MNIST     &   C-2    &  28      &  28  &   0.96 (Acc.)  \\
FashionMNIST  & C-10   &   28    &  28    & 0.71 (Acc.)  \\
Mortality     & C-2 &   76      &  48      &   0.86 (AUC.)  \\
User     &   C-22  &  3          &  50    & 0.61 (Acc.)  \\
Udacity    &   R  &  4096     &  20     & 0.05 (MSE) \\
Energy    &   R  &  22        &  50     & 0.01 (MSE) \\
\bottomrule
\end{tabular}
}

% {
% %\renewcommand{\arraystretch}{0.05}
% \begin{tabular}{c| c r r r c}
% \toprule
%     &&    & & \textbf{\# Training} & \textbf{Victim Clean}            \\
%     \textbf{Dataset} &   \textbf{Task}  &   \textbf{$n$} & \textbf{$L$}& \textbf{Data} & \textbf{Performance} \\
% \midrule
% MNIST     &   C-2    &  28      &  28 & 12,000 &   0.96 (Acc.)  \\
% FashionMNIST  & C-10   &   28    &  28   &   60,000 & 0.71 (Acc.)  \\
% Mortality     & C-2 &   76      &  48   &  14,681   &   0.86 (AUC.)  \\
% User     &   C-22  &  3          &  50   &  22,000 & 0.61 (Acc.)  \\
% Udacity    &   R  &  4096     &  20   &  844  & 0.05 (MSE) \\
% Energy    &   R  &  22        &  50   &  315  & 0.01 (MSE) \\
% \bottomrule
% \end{tabular}
% }

    %\end{adjustbox}
    \caption{Summary of datasets. ``C-N'' means N-class classification, and ``R''means Regression. ``n'' is a  dimension of $x_i$. }
    \label{table-dataset-summary}
    %\vspace{-1.0\baselineskip}
\end{table}
\subsection{Incorporating Different Objectives}
The general form of the framework~(\autoref{eq:problem_definition_attack_loss_general}) allows various attacks through the choice of $\mathrm{Agg}(\cdot)$ and the constraint $\Delta$. 
To showcase our framework's versatility, we choose $\gamma_i$-weighted sum as an instance of Agg($\cdot$):  
\begin{gather}
    \boldsymbol{\delta}=\argmin_{\|{\delta_i}\|_{p} \leq \epsilon,\;\forall i} \sum^{L}_{i=1} \gamma_{i} \mathcal{L}^{\text{adv}}(x_{i},y\in\{y_{i}, y^{a}_{i}\},\delta_{i},h^{\delta}_{i}).
    \label{eq:incorporating_different_objectives}
\end{gather}
%We do not distinguish the current and future terms for brevity.
%We can explain various objectives using $\gamma_{i}$.
In the following, we present three example attacks possible with this aggregation.
\paragraph{ Real-time Attack.} The real-time attack \cite{realtime_adversarial_attack} is a special case of this formulation when $\gamma_{i}= 0$ for $i<L$ and $\gamma_{L}=1$, which aims to mislead the last victim output.
% with temporal specificity.% called the Time-window Attack and Surprise Attack.%, whose purpose is to control the timing of output errors.
\paragraph{Time-window Attack.}
This attack causes misclassification/prediction at only at specific times interval $[a, b]$. This can be useful when 1) the attack has a more impact at specific times, or 2) the attacker has to avoid detection for a time interval where the victim is vigilant. We can implement this attack by setting $y=y^{a}_i, \gamma_i=1$ if $i\in[a,b]$ and $y=y_i, \gamma_i=\tau (>0)$ otherwise in~\autoref{eq:incorporating_different_objectives}.
\paragraph{ Surprise Attack.} Surprise Attack induces untargeted error abruptly by maximizing the difference between the maximum error and the mean error over time: 
\begin{gather}
\begin{aligned}
\argmin_{\|{\delta_i}||_{p} \leq \epsilon,\;\forall i} &\left[  \frac{1}{L}\sum_{i}\mathcal{L}^{\text{adv}}(x_{i},y_{i},\delta_{i},h^{\delta}_{i})\right.\\
 &\left. - \max_j   \mathcal{L}^{\text{adv}}(x_{j},y_{j},\delta_{j},h^{\delta}_{j}) \right].
\end{aligned}
\end{gather}
\if0
$\mathcal{L}_{\text{adv}}(x_{j},y_{j},\delta_{j},h^{\delta}_{j}) - \frac{1}{L-1}\sum_{i\in[1, j-1]\cup[j+1, L]}\mathcal{L}_{\text{adv}}(x_{i},y_{i},\delta_{i},h^{\delta}_{i})$, such that 

$j =\argmax_{j}\mathcal{L}_{\text{adv}}(x_{j},y_{j},\delta_{j},h^{\delta}_{j})$
\fi
%That is, the error at one time point will be particularly large, whereas the errors at another time point remain relatively small, 
This attack prevents a victim from reacting properly, thus causing more damage with the same error. For example, an abrupt steering angle change will be more damaging to an autonomous vehicle than a smooth angle change over time.

\section{Experiment}
We evaluate our attacks to answer the following research questions.
\textbf{RQ1.} How much does Predictive Attack improve the attack performance?,
\textbf{RQ2.} How versatile is our online evasion attack framework?
\textbf{RQ3.} How robust is Predictive Attack?
%\textbf{RQ4.} How efficient is the predictive attack?
% We implemented our attacks using PyTorch on a server equipped with four Titan XP GPUs.
\paragraph{Datasets.} We use six datasets for classification and regression in our evaluations as summarized in \autoref{table-dataset-summary}.
\begin{itemize}
\item \textbf{MNIST}~\cite{mnist}: Given a column sequence (vertical lines) of a digit image, predict the correct label of each column. Two classes, 3 and 8, are selected.
\item \textbf{FashionMNIST}~\cite{fashion-mnist}: The same format as MNIST  but containing clothing images. We use 10 classes for a harder classification problem.
\item \textbf{Mortality}~\cite{mimic_mortality}: Given a sequential medical record, predict a patient's mortality every hour.
\item \textbf{User}~\cite{user-dataset}: Given a sequence of x-y-z accelerations from a user, predict the user of the sequence.
\item \textbf{Udacity}~\cite{udacity}: Given a sequence of camera images, predict steering angles. We resized the images to 64 x 64 and sequence duration is 0.67s.
\item \textbf{Energy}~\cite{energy-dataset}: Given a sequence of 27 weather sensors, predict electricity consumption of a building.
\end{itemize}
\begin{figure}
\centering
\includegraphics[width=1.0\linewidth]{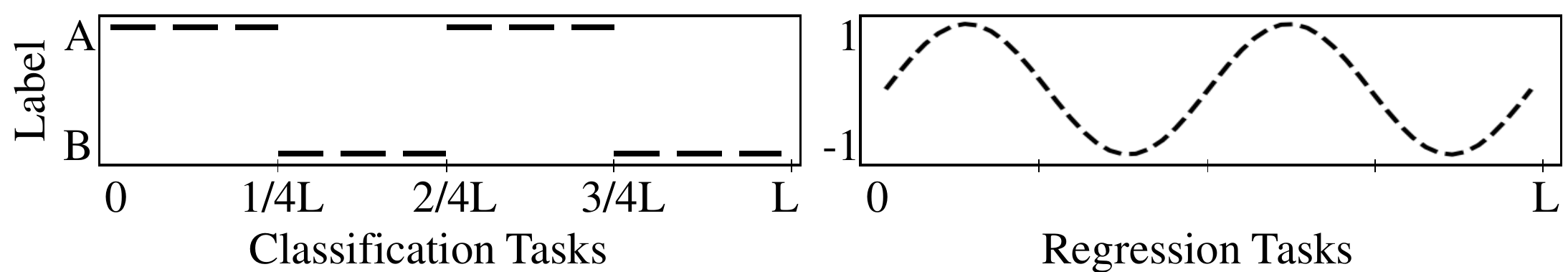}
%\vspace{-1.0\baselineskip}
  \caption{Target label and values of attacks.}
  \label{fig:target label}
\end{figure}
\begin{figure}
\centering
\includegraphics[width=1.0\linewidth]{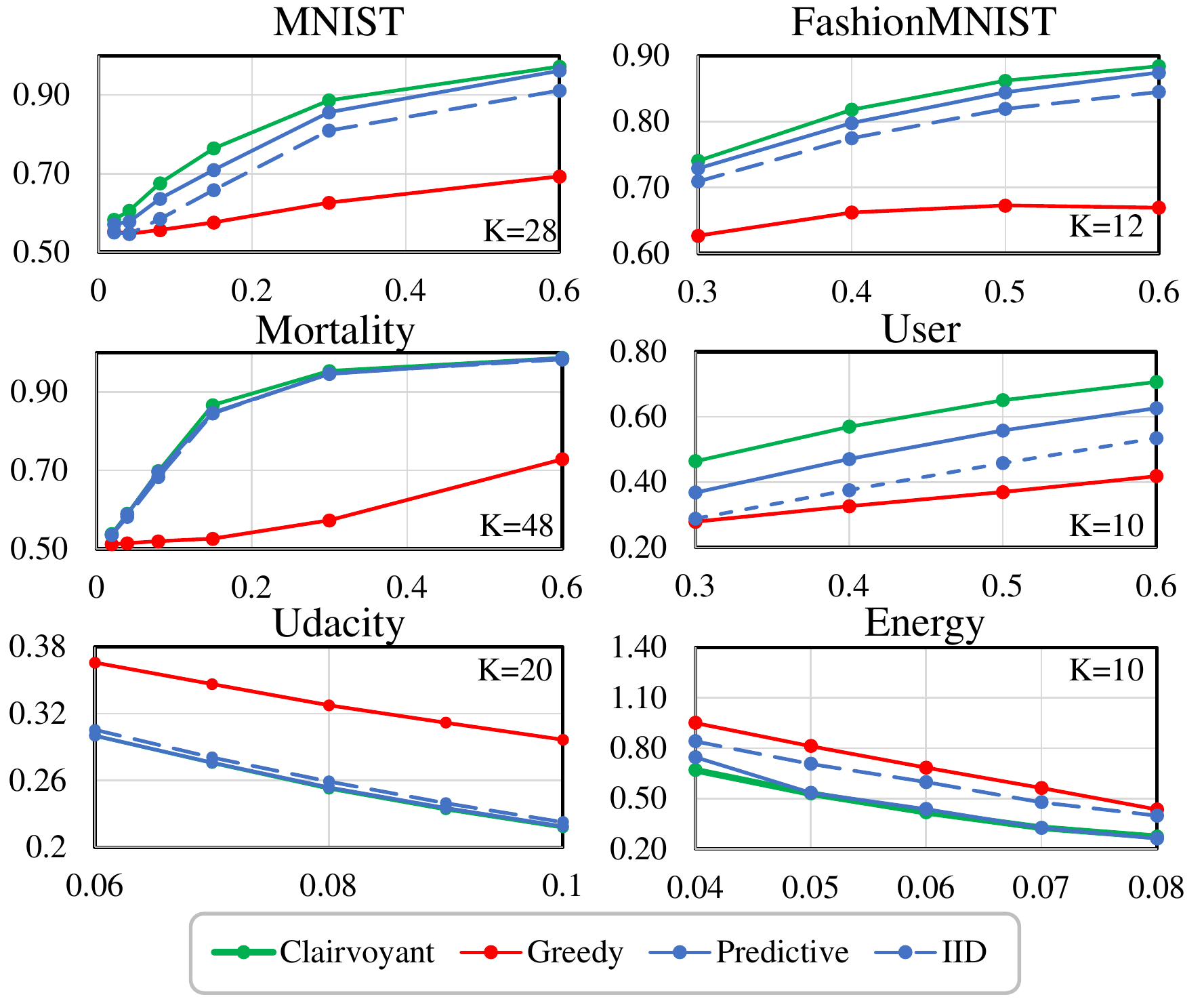}
  \caption{Performance evaluation of Predictive Attack and baselines.}
  \label{fig:performance-evaluation-predictive}
\end{figure}
\begin{figure}
\centering
\includegraphics[width=0.9\linewidth]{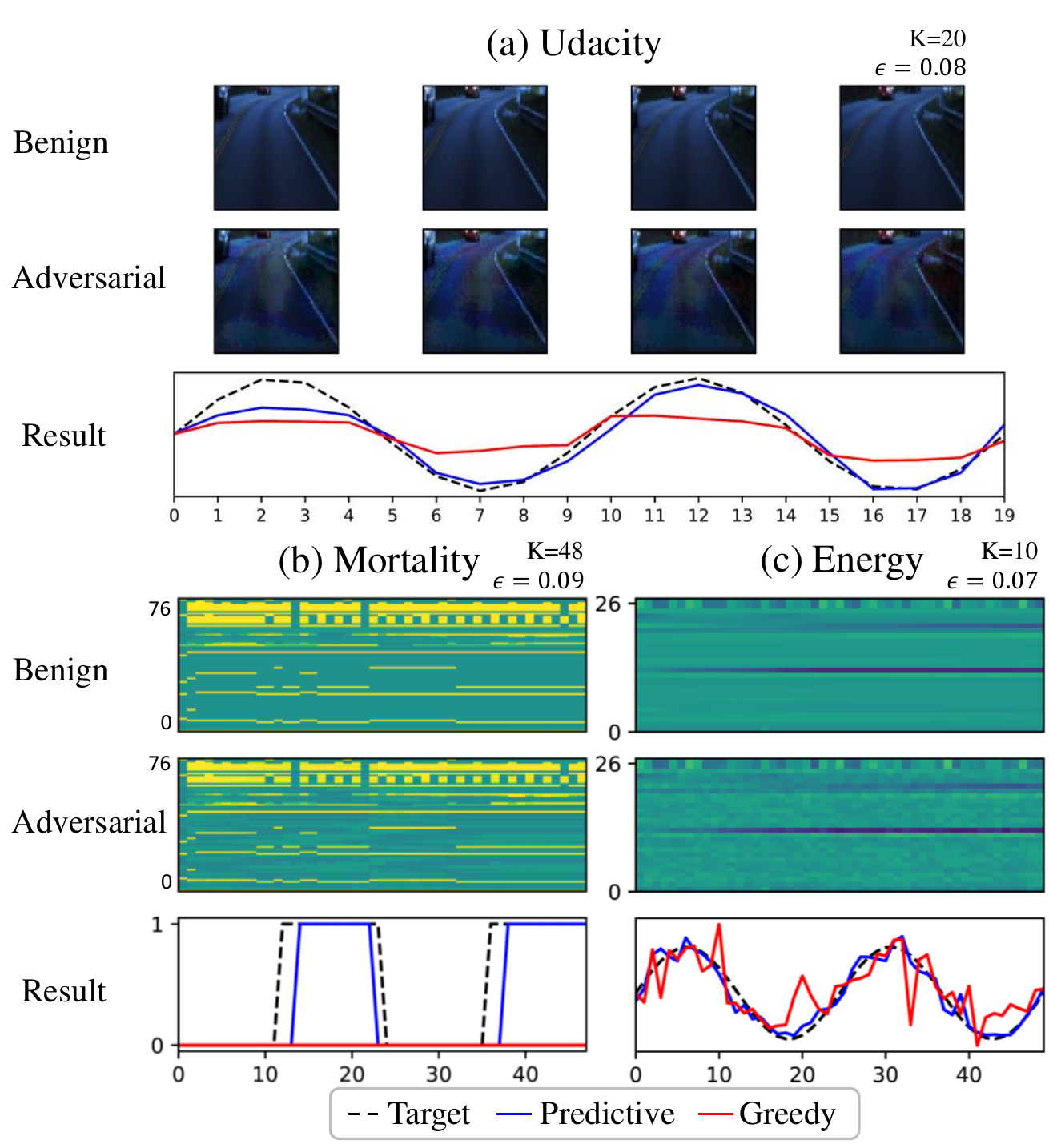}
%\vspace{-0.2\baselineskip}
  \caption{Visualization results of Predictive Attack. We can see Predictive Attack can follow the targeted labels and values much better than Greedy Attack.}
  \label{fig:attack-visualization}
 % \vspace{-0.8\baselineskip}
\end{figure}
\paragraph{Model Parameters.}
%Victim classifiers and future predictors $Q_\phi$ are based on LSTM. %\cite{lstm}.
%In particular, 
All models, except for Udacity, consist of one LSTM layer followed by two linear layers with ReLU activations. For Udacity, we use CNN-LSTM as a victim model, and CrevNet \cite{crevnet} as $Q_\phi$ to deal with the high-dimensional images. More model details are in Appendix B.%~\autoref{appendix-model-parameters}.
We the Adam optimizer for training with a learning rate of 1e-4.
\autoref{table-dataset-summary} summarizes victim's clean performances. 
We use ROC-AUC for Mortality to be comparable to the original reports \cite{mimic_mortality}.
\paragraph{ Attack Target and Performance Metric.}
To evaluate the proposed framework and the attack, we use time-varying target outputs for both classification and regression as depicted in \autoref{fig:target label}.
%Unlike the simple static target output, these targets can 
It is intended to simulate the dynamic nature of real online attacks better. Appendix F contains more results with other target patterns. %\autoref{appendix:different-targets}. 
An effective attack should achieve high TASR and low TMSE. TASR (Targeted Attack Success Ratio) is the number of time steps where a victim model yields targeted labels, over the total number of time steps $L$. TMSE (Targeted Mean Squared Error) is the mean squared error between victim model outputs and the targeted values. We use temporal summation as an attack objective (\autoref{eq:problem_definition_attack_loss}) if not specified.
\paragraph{ Miscellaneous.}
%When we aggregate the losses, we only consider the losses in the last $L/4$ time steps of each sequence instead of all time steps. This is because the victim classifier is initially inaccurate in the beginning of the sequence due to insufficient observation, making any attack too easy in the early phase.
%Under this condition,
%Clairvoyant Attack and Predictive/IID/Random Attack are able to perturb in the interval $[3/4L-K,\;L]$ for a given lookahead $K$, and Greedy Attack can perturb in the interval $[3/4L,L]$ since $K=0$ for Greedy.
%We set $K$ to about 30$\sim$40\% of $L$.
The input values range from 0 to 1, and we use $\ell_\infty$ norm constraints for all tests.
We set $\text{MAX\_ITERS}=100$, and $\alpha=1.5\epsilon /$MAX\_ITERS. We report average results of three experiment repetitions retraining a victim model initialized with random weights. Predicted inputs and the perturbed inputs of our attack are presented %for visual inspection
in Appendices D and E.%Appendices \ref{appendix:prediction-performance} and \ref{appendix:adversarial-examples}.
%We sampled the future $x_i$ once for computing expected loss since it is enough to reach high performances in practice. 
% \begin{itemize}
%     %\item Implementation and overview of experiments
%     %\item Baselines
%     %\item dataset (MNIST, mortality, IMDB sentiment analysis, Udacity self driving)
%     \item Parameter settings
%     \item eps-error 그래프 (epsilon 변화 포함)
%     \item eps-error-k
%     \item eps-error-(MD, IID)
%     \item iter-error
%     \item ranged-attack
%     \item surprise-atack
%     \item vs. realtime
%     %\item defense가 적용된 (예: adversarial training) 모델에 대한 공격
% \end{itemize}

\subsection{Performance Evaluation}
In Figure \ref{fig:performance-evaluation-predictive}, we answer \textbf{RQ1} by comparing the performance of Predictive Attack with Greedy and Clairvoyant. The x-axis is $\epsilon$, $\ell_\infty$ norm of a perturbation, and the y-axis is the performance metric. 
On average at maximum $\epsilon$ of each plot, the performance of Predictive Attack (straight blue) approaches 98\% of Clairvoyant Attacks's TASR (green). Predictive Attack also performs 138\% of Greedy Attack's TASR (red).
%These results show that attempts to predict future inputs and mimic clairvoyants can help overcome online constraints.
In particular, it is worth noting that safety-critical tasks such as Mortality and Udacity are more prone to attacks than the toy datasets, MNIST and Fashion MNIST.

IID Attack's comparable performance (93\% of Predictive Attack on average) to that of Predictive Attack shows the importance of victim model dynamics.
%IID Attack predicts the future by random permutation of the observed input sequence. That is, IID Attack does not use temporal dependence.
%Thus, its performance gain (93\% of Predictive Attack on average) comes from the victim model dynamics.
In particular, in Mortality, IID Attack shows the closest performance to Predictive Attack.
We surmise the victim model is more dependent on model dynamics to solve the mortality prediction task.
For example, a patient's current severity may depend on a medical record several hours ago, not on the current medical record.

% We can see the importance of victim model dynamics for an attack by observing IID Attack's comparable performance to that of Predictive Attack on multiple datasets.
% Since IID Attack predicts the future by random permutation of the observed input sequence,
% its performance improvement only comes from the victim model dynamics, not temporal dependence.
% Only using victim model dynamics, IID Attack performs comparably to Predictive Attack (93\% on average).
% In particular, in Mortality dataset, IID Attack shows the closest performance to predictive.
% We surmise it is because the victim model is more dependent on model dynamics compared with other tasks to solve the mortality prediction task.
% For example, a patient's current severity may depend on a medical record several hours ago, not on the current medical record.

For a qualitative analysis, we visualize the results in \autoref{fig:attack-visualization}. We chose (a) Udacity, (b) Mortality, and (c) Energy for visualization. In each figure section, the three rows correspond to benign examples, corresponding adversarial examples, and victim model outputs, respectively. The x-axis is time. We can see that Predictive Attack (blue) closely follows the target values or labels (dashed black) much better than Greedy Attack (red) can.

\begin{figure}
\centering
\includegraphics[width=0.95\linewidth]{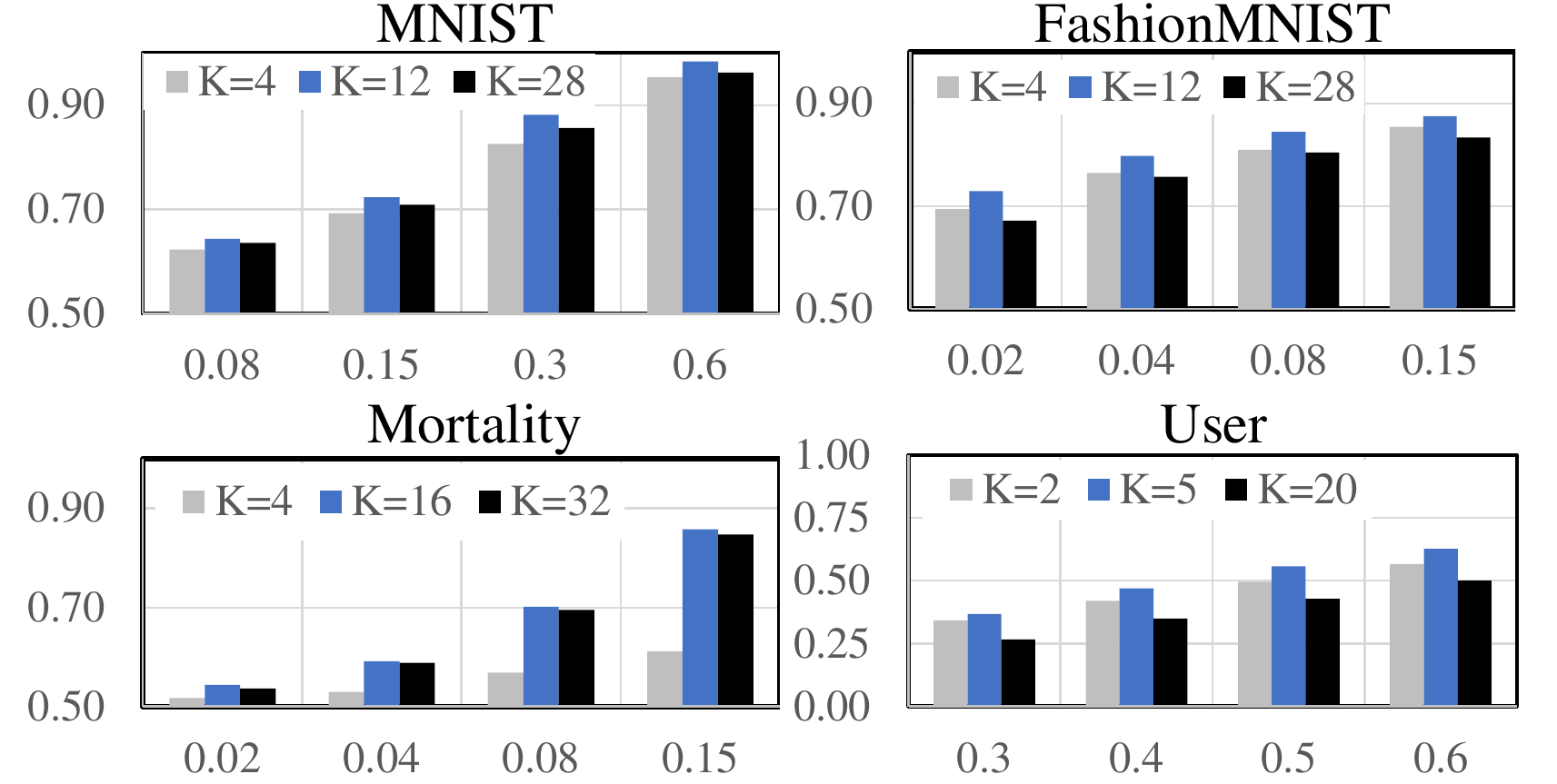}
  %\vspace{-0.6\baselineskip}
  \caption{Effect of the lookahead $K$ on Predictive Attack.}
  %\vspace{-0.9\baselineskip}
  \label{fig:performance-evaluation-k}
\end{figure}

In Figure \ref{fig:performance-evaluation-k}, we investigate the effect of lookahead K.
The x-axis is $\epsilon$, and the y-axis is attack performance. The color of a bar represents a different K. As a result, we find that there is an optimal K. We attribute this to the limitation of $Q_{\phi}$.
By increasing K until $Q_{\phi}$ can predict accurately, the attack can use the longer temporal dependence and victim model dynamics, leading to the performance improvement. However, if K exceeds the limit, the attack performance decreases because of incorrect future inputs and their wrong perturbations.
\begin{figure}
\centering
\includegraphics[width=1.00\linewidth]{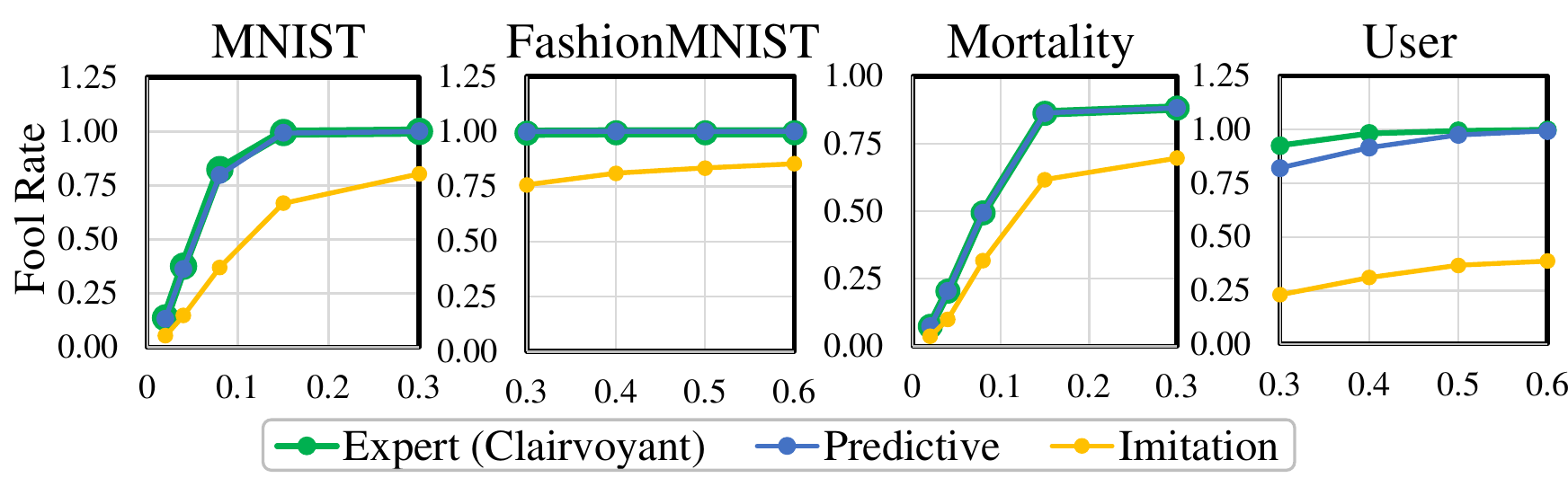}
    \vspace{-1.2\baselineskip}
  \caption{Comparison of attacks with Real-time Attack's objective.} %\cite{realtime_adversarial_attack}
  \label{fig:real-time-attack}
\end{figure}
An attack time should be short to perturb more inputs in a limited time interval. We measured the time per a time step for Predictive Attack to reach 90\% of the saturated performance (when MAX\_ITERATION is used).
For Mortality and Energy, the time is short enough, 0.03 secs and 0.05 secs, considering 3600 secs and 600 secs of each dataset's time step duration. For Udacity, it takes 0.25 secs, which is longer than usual duration of camera input, 0.03 secs. However, we believe that the time can be reduced by using a dedicated hardware or compressing the predictor model.
% An attack should take less time for perturbing more inputs in a limited time interval, which leads to better performance. We measured the time for Predictive Attack to reach 90\% of the saturated performance (when MAX\_ITERATION is used).
% %For small victim models of MNIST and FashionMNIST, Predictive Attack takes 0.02 secs per time step.
% For Mortality and Energy, it takes 0.03 secs and 0.05 secs. The attack speed is fast enough, considering the time step duration of the datasets, respectively 3600 secs and 600 secs. On the other hand, it takes 0.25 secs per time step to attack Udacity victim model. It is not satisfactory since the usual period of camera input is 0.03 secs. However, the current implementation is not optimized, and we believe that the time can be substantially reduced by using a dedicated hardware~\cite{??} or compressing the predictor model~\cite{deep_compression}.
\subsection{Versatility of the Attack Framework}
Evaluating the effectiveness of different objectives from our framework, we answer \textbf{RQ2}, versatility of our framework. %Please refer to the section containing \autoref{eq:incorporating_different_objectives} for a detailed explanation and motivations about the attacks.
\paragraph{ Real-time Attack.}
In Figure \ref{fig:real-time-attack}, we test a real-time attack objective~\cite{realtime_adversarial_attack} as a special case of our framework. A real-time attacker aims to mislead the last output of a victim model and is untargeted; thus, y-axis is ``Fool Rate'' that means a frequency of wrong victim decisions.
Imitation learning-based real-time attack~ \cite{realtime_adversarial_attack} is reimplemented, referring to the public codes\footnote{\url{https://github.com/YuanGongND/realtime-adversarial-attack}} (See 
Appendix K for detail explanations.).
%\autoref{appendix:real-time-attack-validation} for detail explanations.).
Predictive Attack surpasses the imitation learning-based attack (88$\sim$100\% vs. 24$\sim$94\% of experts' performance).
Note that this gain of Predictive Attack comes at the cost of solving PGD, unlike the imitation learning-based attack that depends on a pre-trained agent.
Predictive Attack is good for achieving a high attack performance, while imitation learning-based attack is suitable for fast attacks.
% We test an objective proposed in real-time attack~\cite{realtime_adversarial_attack} as a special case of our framework. The objective is to mislead the last output of a victim model, and is untargeted. 
% We reimplemented imitation learning-based attack introduced by \cite{realtime_adversarial_attack} referring to the public implementation\footnote{\url{https://github.com/YuanGongND/realtime-adversarial-attack}} (See Appendix H for detail explanations.).
% In Figure \ref{fig:real-time-attack}, Predictive Attack surpasses the imitation learning-based attack, consistently over the different datasets.
% Predictive Attack achieve 88$\sim$100 \% of experts' performance. 
% In contrast, The imitation learning-based attack shows about 24$\sim$94\% of experts' performance.
% Note that this gain of Predictive Attack comes at the cost of solving PGD at each time step, unlike the imitation learning-based attack that depends on a pre-trained agent which is less effective to unseen test inputs.
% Predictive Attack is proper for a task where a high attack performance is required, while the imitation learning-based attack is suitable when input arrival frequency is high.
\begin{figure}
\centering
\includegraphics[width=1.0\linewidth]{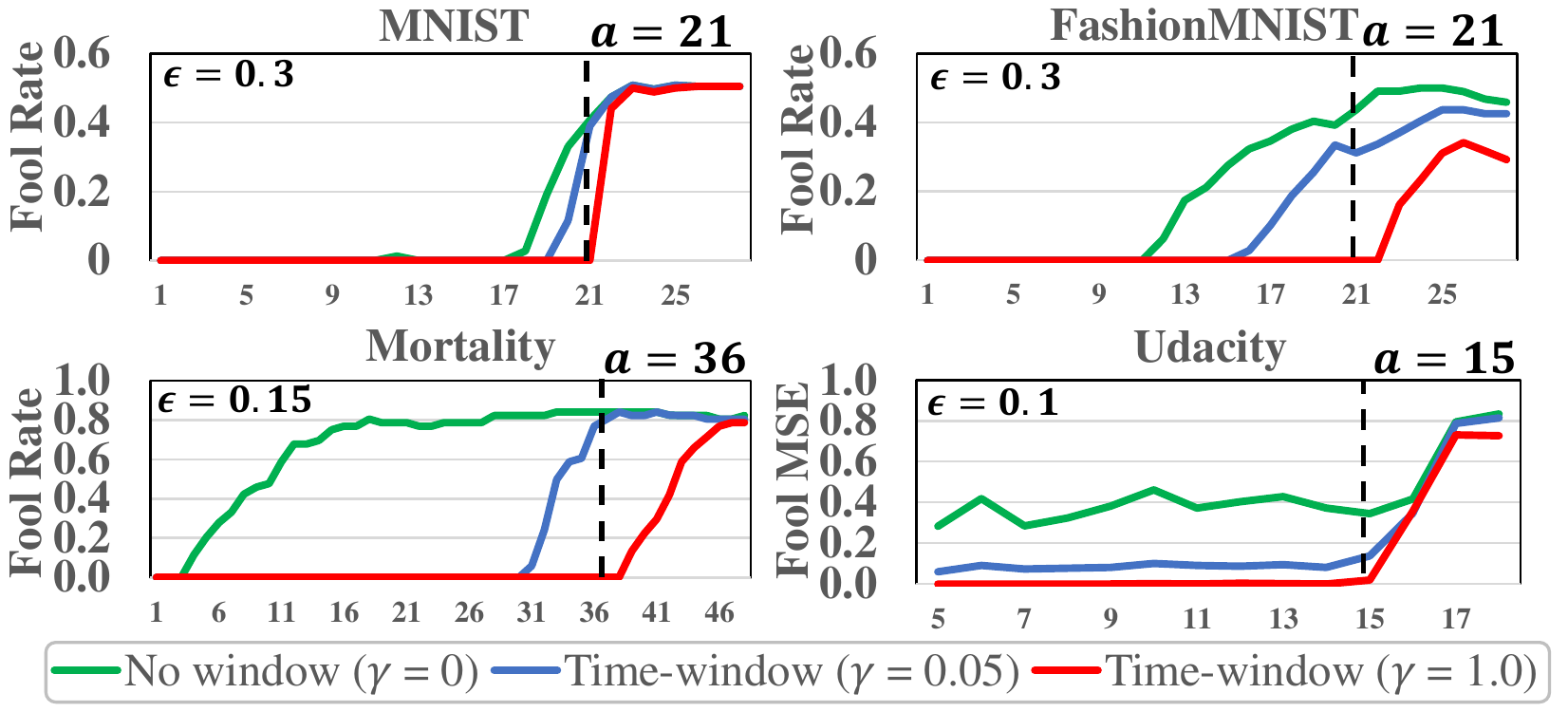}
%\vspace{-1.5\baselineskip}
  \caption{Effect of using the Time-window Attack objective whose purpose is to restrict the error to the interval $[3/4L,L]$.
  Note that non-window attacks cause error before this interval.
  }%\JH{How about we just explain what the experiments were?}
    
  \label{fig:time-constraint-attack}
  %\vspace{-0.5\baselineskip}
\end{figure}
\begin{figure}
\centering
\includegraphics[width=1\linewidth]{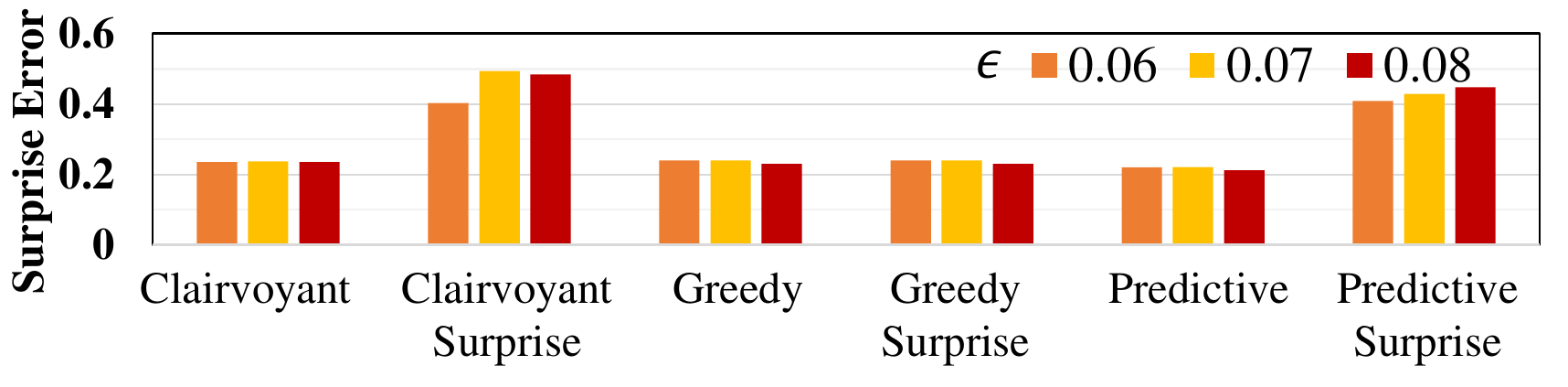}
%\vspace{-2.0\baselineskip}
  \caption{
  %Our attack formulation can be configured to accommodate Surprise Attack's objective.
  Effect of using the Surprise Attack objective. It aims  to cause a sudden error and disrupt the victim from responding properly. 
  %\JH{How about we just explain what the experiments were?}
  }
  \label{fig:surprise-attack}
  %\vspace{-0.5\baselineskip}
\end{figure}
\paragraph{ Time-window Attack.}
%We demonstrate the versatility of our attack formulation with Time-window Attack. 
We demonstrate the attack with temporal specificity. 
We chose the interval $[a=3/4L, b=L]$ as the intended window of error. 
In Figure \ref{fig:time-constraint-attack}, we present the performance of Predictive Attack with/without the Time-window objective. The x-axis is time. We set the y-axis as ``Fool Rate'' and ``Fool MSE'' (MSE between true values and victim outputs)
%since TASR and TMSE can not demonstrate temporal specificity
. An ideal attack should cause non-zero values only in $[3/4L, L]$.
Predictive Attack %with Time-window 
fulfills this objective and increases the error after $t$=$a$ in contrast to non-window attacks.
We also find $\tau$ controls the trade-off between attack performance and compliance with the time-window. 
\paragraph{ Surprise Attack.}
We conduct Surprise Attack experiment with the autonomous driving task from Udacity, where Surprise Attack can be practically important. We define Surprise Error as $\text{max}_{i} |y_i - f(x_i, h^\delta_i)| - \text{mean}_{i} |y_i - f(x_i, h^\delta_i)|$.
In \autoref{fig:surprise-attack}, Predictive Attack with Surprise objectives achieves about 2.09 times higher Surprise Error than a naive Predictive Attack and Greedy at $\epsilon = 0.08$. 

\begin{table}[t]
    \centering
    %\scriptsize
    \footnotesize
    %\small
    %\normalsize
    %\begin{adjustbox}{minipage=\linewidth,scale=1.0}
    %\centering
    {
\begin{tabular}{c|r r| r r |r}
\toprule
& & &\multicolumn{2}{c|}{\textbf{Predictive}} & \textbf{Greedy}\\
    \textbf{Dataset} &   \textbf{$\epsilon$}  &   K & \textbf{$\eta=0$}& \textbf{$\eta=0.4$}&  \\
    
\midrule
MNIST     &   0.08    &  8   &   0.66 &   0.64 & 0.56  \\
FashionMNIST  & 0.30   &   8    &  0.76   &   0.74 & 0.63  \\
Mortality     & 0.15 &   32  &  0.85   &   0.80 & 0.52  \\
User     &   0.30  &  10 &   0.34 & 0.25 & 0.28  \\
Udacity (MSE)     &   0.05  &  16 &   0.35 & 0.37 & 0.41 \\
%Energy  (MSE)   &   0.15  &  20 &   0.21 & 0.35 & 0.25  \\
\bottomrule
\end{tabular}
}

    \caption{Predictive Attack against incorrect future prediction.}
    %\vspace{-0.8\baselineskip}
    %\end{adjustbox}
    \label{table:noise-robustness}
\end{table}

\begin{table}[t]
    \centering

    %\scriptsize
    \footnotesize
    %\small
    %\begin{adjustbox}{minipage=\linewidth,scale=1.0}
      \centering
      {
\begin{tabular}{c|r r| r r}
\toprule
    \textbf{Dataset} &   \textbf{$\epsilon$}  &   K & \textbf{Whitebox}& \textbf{Graybox} \\
\midrule
MNIST     &   0.3    &  28   &   0.86 &   0.64  \\
FashionMNIST  & 0.5   &   28    &  0.88   &   0.47  \\
Mortality     & 0.15 &   32  &  0.85   &   0.75  \\
User     &   0.5  &  10 &   0.54 & 0.21  \\
Udacity (MSE)    &   0.06  &  16 &   0.30  & 0.47 \\
\bottomrule
\end{tabular}
}

    %\end{adjustbox}
        \caption{Predictive Attack when model parameters are unknown.}
    %\vspace{-1.6\baselineskip}
    \label{table:blackbox-transfer}
\end{table}

\subsection{Robustness Evaluation}
To answer \textbf{RQ3}, we evaluate Predictive Attack under a variety of unseen situations in our attack framework.
\paragraph{ Incorrect Future Prediction.}
We investigate the performance of Predictive Attack under degraded $Q_\phi$.
To control the prediction quality, we replace a predicted future input $x_t$ with  $x^{\eta}_{i}=(1-\eta)x_{i}+\eta e$, where $e$ is a uniform random variable in the valid input range.
In \autoref{table:noise-robustness},
although slightly decreased as the noise is added ($\eta=0.4$), Predictive Attack performs better than Greedy Attack, using the victim model dynamic as consist with the case of IID Attack in  \autoref{fig:performance-evaluation-predictive}.
\paragraph{ Unknown Model Parameters.}
To evaluate the robustness under limited victim information, 
a transfer attack~\cite{delving} is conducted (\autoref{table:blackbox-transfer}). We assume a gray-box threat model where an attacker knows a victim model's architecture but not model parameters. Adversarial examples generated from an attacker-trained surrogate model  are transferred to the actual victim model.
Transfer attack achieves average 63\% performance of white-box attack in the classification tasks, up to 88\% in Mortality.

\section{Conclusion}
This paper introduces a general framework for online evasion attacks on recurrent models. Our framework can accommodate various time-varying attack objectives and constraints, allowing a comprehensive robustness analysis. 
Based on our framework, we propose Predictive Attack and IID Attack. The success of these attacks highlights the new surface of attack for recurrent models, which need to be addressed.
However, defense in the online setting has not been fully studied yet, while existing offline defenses~\cite{pgd,trade} are not suitable for online tasks. 
We leave it as future work to investigate online defense methods.

\newpage
\section*{Acknowledgements}
This work was supported in part by  ERC (NRF-2018R1A5A1059921) and IITP (2020-0-00209) funded by the Korea government(MSIT). It is also supported in part by the NSF EPSCoR-Louisiana Materials
Design Alliance (LAMDA) program \#OIA-1946231.
\bibliographystyle{named}
%\bibliography{ijcai22}
\bibliography{acmart}

% \newpage
% \begin{appendices}
% \appendix
% \input{0_IJCAI22/appendix}
% \end{appendices}
\end{document}